\documentclass{iaus}
\usepackage{graphicx}

\title[Convection and $^6$Li in metal-poor halo stars] %% give here short title %%
{Convection and $^6$Li in the atmospheres of metal-poor halo stars}

\author[M. Steffen et al.]   %% give here short author list %%
{M. Steffen$^1$,
R. Cayrel$^2$, P. Bonifacio$^2$, H.-G. Ludwig$^3$, E. Caffau$^2$}

\affiliation{$^1$Astrophysiklaisches Institut Potsdam, An der Sternwarte 16,
D-14482 Potsdam, Germany 
\\[\affilskip]
$^2$GEPI -- Observatoire de Paris, Paris, France \\[\affilskip]
$^3$ZAH-Landessternwarte, K\"onigstuhl 12, D-69117 Heidelberg, Germany}

\pubyear{2010}
\volume{268}  %% insert here IAU Symposium No.
\jname{Light elements in the Universe}
\editors{C. Charbonnel, M. Tosi, F. Primas \& C. Chiappini, eds.}
\begin{document}

\maketitle

\begin{abstract}
Based on 3D hydrodynamical model atmospheres computed with the
CO$^5$BOLD code and 3D non-LTE (NLTE) line formation calculations, 
we study the effect of the convection-induced line asymmetry on the 
derived $^6$Li abundance for a range in effective temperature, gravity,
and metallicity covering the stars of the 
\cite[Asplund \etal\ (2006)]{A2006} sample.
When the asymmetry effect is taken into account for
this sample of stars, the resulting $^6$Li/$^7$Li ratios are reduced by 
about 1.5\% on average with respect to the isotopic ratios determined
by \cite[Asplund \etal\ (2006)]{A2006}. This purely theoretical correction 
diminishes the number of significant $^6$Li detections from 9 to 4 
(2$\sigma$ criterion), or from  5 to 2 (3$\sigma$ criterion). In view of 
this result the existence of a $^6$Li plateau appears questionable.
A careful reanalysis of individual objects by fitting the observed 
lithium $6707$~\AA\ doublet both with 3D~NLTE and 1D~LTE synthetic 
line profiles confirms that the inferred $^6$Li abundance 
is systematically lower when using 3D NLTE instead of 1D LTE line fitting.
Nevertheless, halo stars with unquestionable $^6$Li detection do exist
even if analyzed in 3D-NLTE, the most prominent example being HD\,84937.
\end{abstract}

\firstsection % if your document starts with a section,
\section{Introduction}
The spectroscopic signature of the presence of $^6$Li in the
atmospheres of metal-poor halo stars is a subtle extra depression in
the red wing of the $^7$Li doublet, which can only be detected in
spectra of the highest quality. Based on high-resolution, high
signal-to-noise VLT/UVES spectra of 24 bright metal-poor stars,
\cite[Asplund \etal\ (2006)]{A2006} report the detection of $^6$Li
in nine of these objects. The average $^6$Li/$^7$Li isotopic ratio in
the nine stars in which $^6$Li has been detected is about 4\% 
and is very similar in each of these stars, defining a $^6$Li plateau at
approximately $\log n(^6$Li$) = 0.85$ (on the scale $\log n($H$) = 12$).
A convincing theoretical explanation of this new $^6$Li plateau
turned out to be problematic: the high abundances of $^6$Li at the 
lowest metallicities cannot be explained by current models of galactic 
cosmic-ray production, even if the depletion of $^6$Li during the 
pre-main-sequence phase is ignored (see reviews by e.g.\ 
\cite[Christlieb 2008]{Ch2008}, \cite[Cayrel \etal\ 2008]{Ca2008}, 
Prantzos 2010 [this volume] and references therein).

A possible solution of the so-called `second Lithium problem' was
proposed by \cite[Cayrel \etal\ (2007)]{Ca2007}, who point out that
the intrinsic line asymmetry caused by convection in the photospheres
of metal-poor turn-off stars is almost indistinguishable from the 
asymmetry produced by a weak $^6$Li blend on a presumed symmetric 
$^7$Li profile. As a consequence, the derived $^6$Li abundance should 
be significantly reduced when the intrinsic line asymmetry in properly 
taken into account. Using 3D NLTE line formation calculations based 
on 3D hydrodynamical model atmospheres computed with the CO$^5$BOLD code
(\cite[Freytag \etal\ 2002]{F2002}, \cite[Wedemeyer \etal\
2004]{W2004}, see also
{\tt \small http://www.astro.uu.se/$\sim$bf/co5bold\_main.html}),
we quantify the theoretical effect of the convection-induced line
asymmetry on the resulting $^6$Li abundance as a function of effective
temperature, gravity, and metallicity, for a parameter range that covers
the stars of the \cite[Asplund \etal\ (2006)]{A2006} sample.

A careful reanalysis of individual objects is under way,
in which we consider two alternative approaches for 
fixing the residual line broadening, $V_{\rm BR}$, the combined 
effect of macroturbulence (1D only) and instrumental broadening, for
given microturbulence (1D only) and rotational velocity: 
(i) treating $V_{\rm BR}$ as a free parameter when fitting the Li feature, 
(ii) deriving $V_{\rm BR}$ from additional unblended spectral lines with 
similar properties as Li\,I\,$6707$. We show that method (ii) is potentially 
dangerous, because the inferred broadening parameter shows considerable 
line-to-line variations, and the resulting $^6$Li abundance depends rather 
sensitively on the adopted value of $V_{\rm BR}$.\\[-6mm]

\section{3D hydrodynamical simulations and spectrum synthesis}
The hydrodynamical atmospheres used in the present study are part of the
CIFIST 3D model atmosphere grid (\cite[Ludwig \etal\ 2009]{L2009}). They have
been obtained from realistic numerical simulations with the CO$^5$BOLD code
which solves the time-dependent equations of compressible hydrodynamics in
a constant gravity field together with the equations of non-local,
frequency-dependent radiative transfer in a Cartesian box representative of
a volume located at the stellar surface. The computational domain is periodic
in $x$ and $y$ direction, has open top and bottom boundaries, and is resolved
by typically 140$\times$140$\times$150 grid cells. The vertical optical depth
of the box varies from $\log \tau_{\rm Ross} \approx -8$ (top) to
$\log \tau_{\rm Ross} \approx +7.5$ (bottom), and the radiative transfer is 
solved in 6 or 12 opacity bins. Further information about the  models 
used in the present study is compiled in Table\,\ref{tab1}. Each of the models 
is represented by a number of snapshots, indicated in column (6), chosen from
the full time sequence of the corresponding simulation. 
 
\begin{figure}
%\vspace*{-2.0 cm}
\begin{center}
\mbox{\includegraphics*[bb=44 38 560 390,width=0.96\textwidth]
{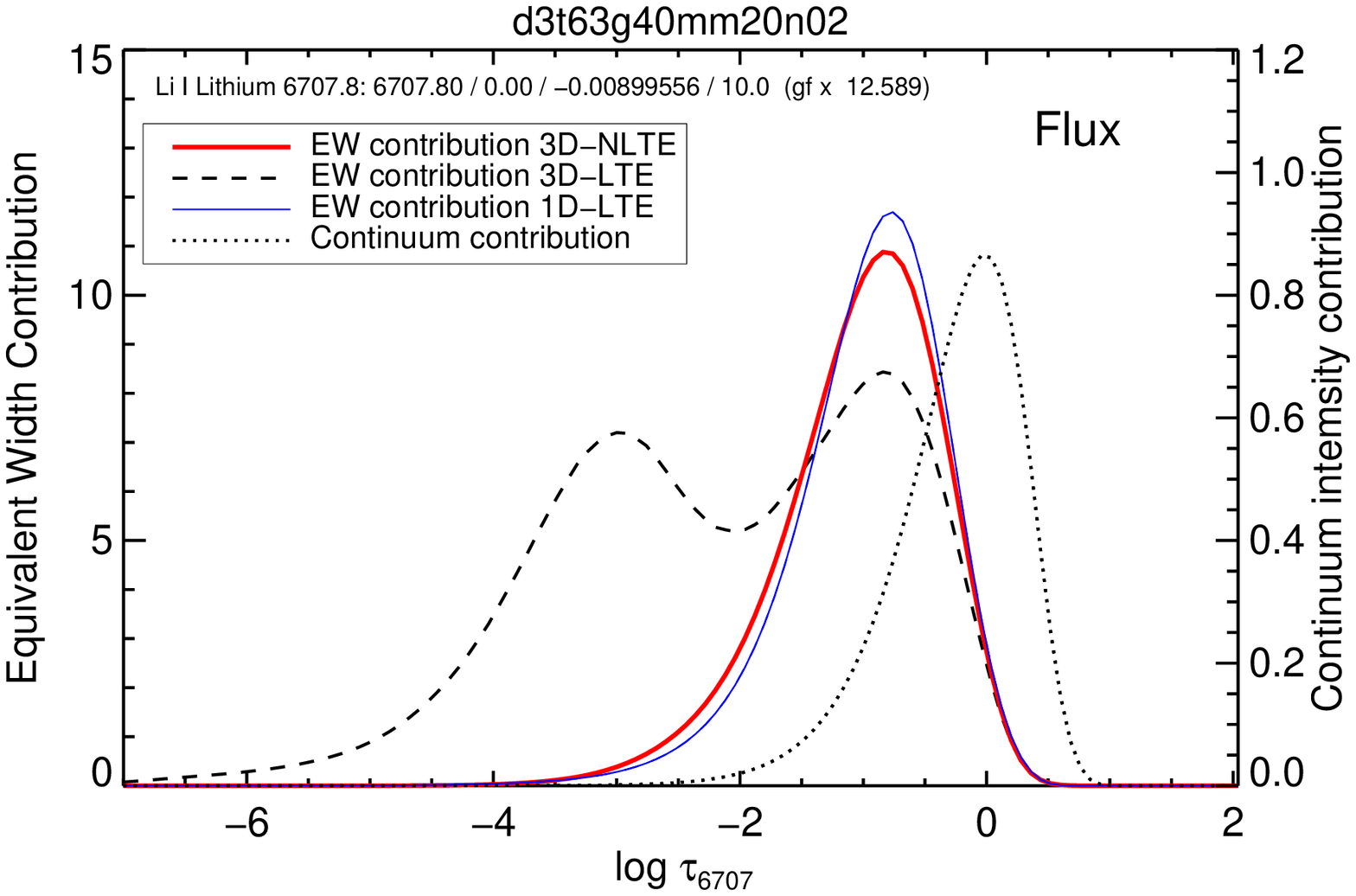}}
\mbox{\includegraphics*[bb=24 14 580 370,width=0.96\textwidth]
{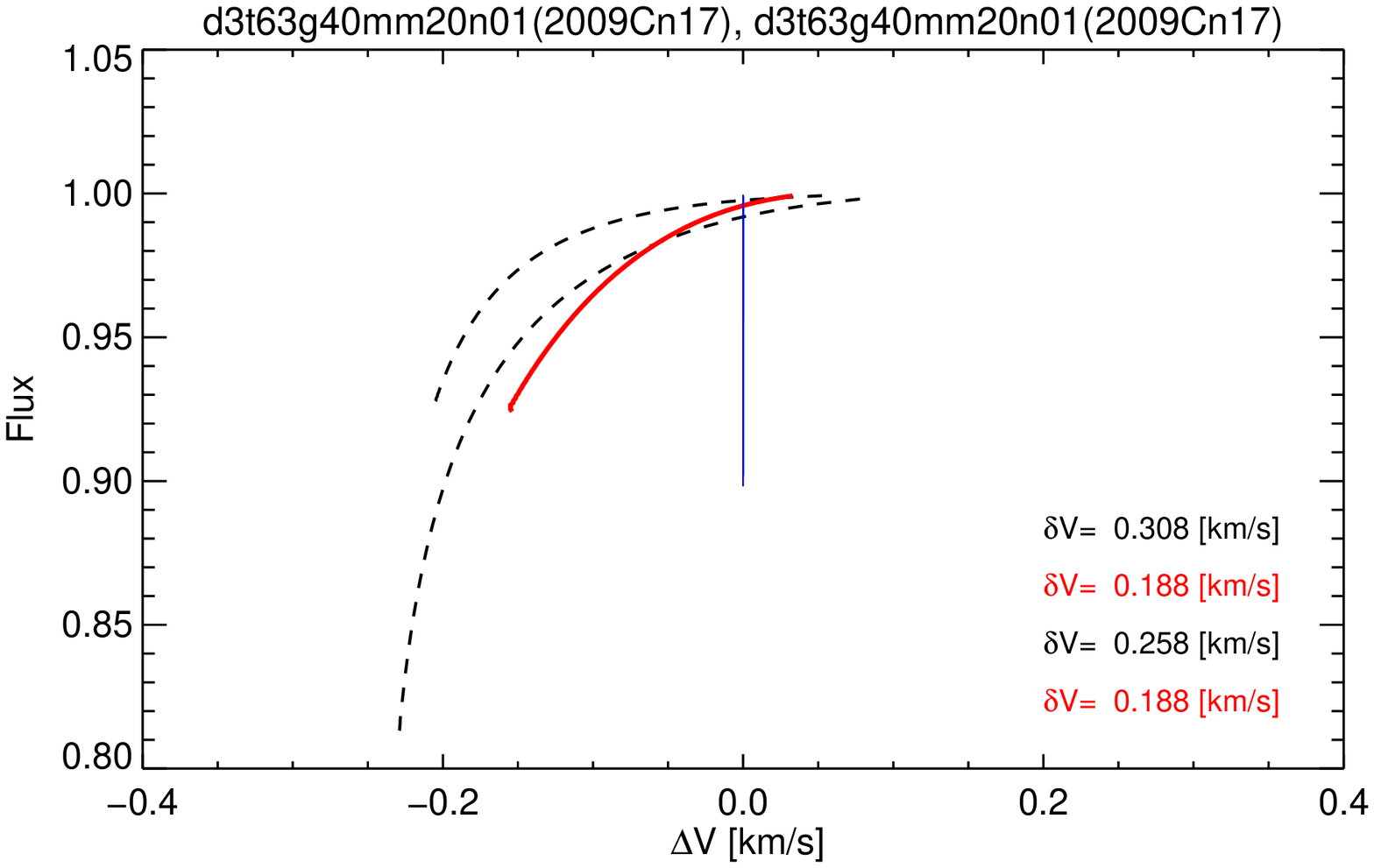}}
\end{center}
% \vspace*{-1.0 cm}
\caption{Comparison of 3D LTE (dashed), 3D non-LTE (thick solid), and 1D LTE
(thin solid) equivalent width contribution functions (top) and line bisectors
(bottom) of a single $^7$Li component, computed for a typical metal-poor 
turn-off halo star ($T_{\rm eff}=6215$~K, $\log g=4.0$, [Fe/H] $=-2$). 
Non-LTE effects strongly reduce the height range of formation and equivalent 
width of the line ($W$: $35.5 \rightarrow  15.6$~m\AA), while differences 
between 1D LTE and 3D NLTE are much smaller. The line asymmetry
is smaller in non-LTE: the velocity span of the bisector is 
$\delta v$(NLTE)~$\approx 190$~m/s, while $\delta v$(LTE)~$\approx 310$ and
$260$~m/s, respectively, before and after adjusting the equivalent width 
(by reducing the Li abundance) and the half-width (by Gaussian broadening)
of the LTE line to match the values of the NLTE profile (compare the 
two dashed bisectors).
}
\label{fig1}
\end{figure}

These representative snapshots are processed by the non-LTE code NLTE3D that 
solves the statistical equilibrium equations for a 17 level lithium atom with 
34 line transitions, fully taking into account the 3D thermal structure of
the respective model atmosphere. The photo-ionizing radiation field is
computed at $704$ frequency points between $\lambda\,925$ and 32\,407~\AA,
using the opacity distribution functions of \cite{CK2004} to allow for
metallicity-dependent line-blanketing, including the H\,I--H$^+$ and 
H\,I--H\,I quasi-molecular absorption near $\lambda\,1400$ and $1600$~\AA, 
respectively. Collisional ionization by neutral hydrogen via the charge 
transfer reaction 
H($1s$) + Li($n\ell$) $\leftrightarrow$ Li$^+$($1s^2$) + H$^-$ is 
treated according to \cite{barklem2003}. More details are given in 
\cite{S2009}. Finally, 3D NLTE synthetic line profiles of the 
Li\,I $\lambda\,6707$~\AA\ doublet are computed with the line formation code 
Linfor3D ({\tt \small http://www.aip.de/$\sim$mst/linfor3D\_main.html}), using
the departure coefficients $b_i$\,=\,$n_i({\rm NLTE})/n_i({\rm LTE})$
provided by NLTE3D for each level $i$ of the lithium model atom as a function
of geometrical position within the 3D model atmospheres. As demonstrated in
Fig.\,\ref{fig1}, 3D NLTE effects are very important for the metal-poor
dwarfs considered here: they strongly reduce the height range of line 
formation such that the 3D NLTE equivalent width is smaller by roughly a 
factor 2 compared to 3D LTE. Ironically, the line strength predicted by 
standard 1D mixing-length models in LTE are close to the results obtained 
from elaborate 3D NLTE calculations. We note that the half-width of the 3D 
NLTE line profile, FWHM(NLTE)\,=\,8.5~km/s, is larger by about 10\%:
FWHM(LTE)\,=\,$7.7$ and $7.5$~km/s, respectively, before and after reducing 
the Li abundance such that 3D LTE and 3D NLTE equivalent widths agree. 
This is because 3D LTE profile senses the higher photosphere where both
thermal and hydrodynamical velocities are lower. However, the NLTE 
line profile is significantly less asymmetric than the LTE
profile, even if the latter is broadened to the same half-width
(Fig.\,\ref{fig1}, bottom panel).\\[-6mm]

\begin{table}[bt]
  \begin{center}
  \caption{List of models used in the present study. Columns (2)-(6) give 
           effective temperature, surface gravity, metallicity, number of 
           opacity bins used in the radiation hydrodynamics simulation, and 
           number of snapshots selected for spectrum synthesis. The equivalent 
           width of the synthetic 3D non-LTE $^7$Li doublet at 
           $\lambda\,6707$~\AA, assuming A(Li)=2.2 and no $^6$Li, 
           is given in column (7). Columns (8) and (9) show $q^\ast_a$(Li) and
           $q^\ast_b$(Li), the $^6$Li/$^7$Li isotopic ratio inferred from 
           fitting this 3D non-LTE line profile with two different kinds of 1D 
           profiles, in each case assuming a rotational broadening of 
           $v\,\sin i$ = 0.0 / 2.0 km/s, respectively (see text for details).}
  \label{tab1}
 {\scriptsize
  \begin{tabular}{|c|c|c|c|c|c|c|c|c|}\hline
{\bf Model}               & {\bf $T_{\rm eff}^{~~1)}$} & {\bf $\log g$}  & 
{\bf [Fe/H]}              & {\bf Nbin}                & {\bf Nsnap}     &
{\bf W$^{ 2)}$}           & {\bf $q^\ast_a$=\,n($^6$Li)/n($^7$Li)} & {\bf $q^\ast_b=\,$n($^6$Li)/n($^7$Li)} \\
{}                        & {\bf [K]}                 &                 & 
                          &                           &                 & 
{[m\AA]}                  & $\langle$3D$\rangle$ NLTE~~~[\%] & 1D LTE~~~[\%]       \\
\hline
d3t59g40mm30n02& $5846$ & $4.0$ & -3.0 & $  6$ & 20 & 44.9 & 1.14 / 1.14 & {\bf 0.88 / 0.88}\\
d3t59g45mm30n01& $5924$ & $4.5$ & -3.0 & $  6$ & 19 & 39.9 & 0.75 / 0.75 & {\bf 0.63 / 0.64}\\
d3t63g40mm30n01& $6269$ & $4.0$ & -3.0 & $  6$ & 20 & 23.9 & 1.96 / 1.93 & {\bf 1.86 / 1.83}\\
d3t63g40mm30n02& $6242$ & $4.0$ & -3.0 & $ 12$ & 20 & 24.1 & 1.81 / 1.80 & {\bf 1.63 / 1.62}\\
d3t63g45mm30n01& $6272$ & $4.5$ & -3.0 & $  6$ & 18 & 24.3 & 1.07 / 1.06 & {\bf 1.02 / 1.00}\\
d3t65g40mm30n01& $6408$ & $4.0$ & -3.0 & $  6$ & 20 & 20.0 & 1.75 / 1.70 & {\bf 1.70 / 1.66}\\
d3t65g45mm30n01& $6556$ & $4.5$ & -3.0 & $  6$ & 12 & 16.4 & 1.29 / 1.27 & {\bf 1.25 / 1.22}\\
\hline
d3t59g35mm20n01& $5861$ & $3.5$ & -2.0 & $  6$ & 20 & 42.8 & 2.48 / 2.44 & {\bf 2.02 / 2.01}\\
d3t59g40mm20n02& $5856$ & $4.0$ & -2.0 & $  6$ & 20 & 45.2 & 1.59 / 1.59 & {\bf 0.96 / 0.98}\\
d3t59g45mm20n01& $5923$ & $4.5$ & -2.0 & $  6$ & 18 & 42.3 & 1.12 / 1.13 & {\bf 0.45 / 0.46}\\
d3t63g35mm20n01& $6287$ & $3.5$ & -2.0 & $  6$ & 20 & 22.1 & 4.09 / 4.00 & {\bf 4.04 / 3.97}\\
d3t63g40mm20n01& $6278$ & $4.0$ & -2.0 & $  6$ & 16 & 23.7 & 1.95 / 1.94 & {\bf 1.79 / 1.78}\\
d3t63g40mm20n02& $6215$ & $4.0$ & -2.0 & $ 12$ & 20 & 25.1 & 1.92 / 1.92 & {\bf 1.66 / 1.67}\\
d3t63g45mm20n01& $6323$ & $4.5$ & -2.0 & $  6$ & 19 & 23.0 & 1.18 / 1.18 & {\bf 0.97 / 0.97}\\
d3t65g40mm20n01& $6534$ & $4.0$ & -2.0 & $  6$ & 19 & 16.2 & 2.28 / 2.21 & {\bf 2.22 / 2.16}\\
d3t65g45mm20n01& $6533$ & $4.5$ & -2.0 & $  6$ & 19 & 17.0 & 1.31 / 1.29 & {\bf 1.21 / 1.19}\\
\hline
d3t59g40mm10n02& $5850$ & $4.0$ & -1.0 & $  6$ & 20 & 41.5 & 1.60 / 1.62 & {\bf 1.45 / 1.47}\\
d3t59g45mm10n01& $5923$ & $4.5$ & -1.0 & $  6$ & 08 & 38.2 & 1.14 / 1.15 & {\bf 0.83 / 0.85}\\
d3t63g40mm10n01& $6261$ & $4.0$ & -1.0 & $  6$ & 20 & 22.0 & 2.37 / 2.38 & {\bf 2.33 / 2.33}\\
d3t63g40mm10n02& $6236$ & $4.0$ & -1.0 & $ 12$ & 20 & 21.8 & 2.15 / 2.16 & {\bf 2.05 / 2.06}\\
d3t63g45mm10n01& $6238$ & $4.5$ & -1.0 & $  6$ & 20 & 23.4 & 1.36 / 1.37 & {\bf 1.23 / 1.24}\\
d3t65g40mm10n01& $6503$ & $4.0$ & -1.0 & $  6$ & 20 & 15.5 & 3.10 / 3.02 & {\bf 3.14 / 3.06}\\
d3t65g45mm10n01& $6456$ & $4.5$ & -1.0 & $  6$ & 19 & 17.1 & 1.51 / 1.51 & {\bf 1.44 / 1.43}\\
\hline
  \end{tabular}
  }
 \end{center}
\vspace{1mm}
 \scriptsize{
 {\it Notes:}
  $^{1)}$ averaged over selected  snapshots; $^{2)}$ averaged over selected 3D non-LTE 
  spectra}
\end{table}

\section{$^6$Li bias due to convective line asymmetry}
As outlined above, the $^6$Li abundance is systematically overestimated if one
ignores the intrinsic asymmetry of the $^7$Li line components. To quantify this
bias theoretically, we rely on synthetic spectra. The idea is as follows: we 
represent the observation by the synthetic 3D NLTE line profile of the $^7$Li 
line blend, computed with zero $^6$Li content. Except for an optional rotational
broadening, the only source of non-thermal line broadening is the 
3D hydrodynamical velocity field, which also gives rise to a convective 
blue-shift and an intrinsic line asymmetry. Now this 3D $^7$Li line blend is 
fitted by 'classical' 1D synthetic line profiles composed of intrinsically 
symmetric components of $^6$Li and $^7$Li. Four parameters are varied 
independently to find the best fit (minimum $\chi^2$): 
in addition to the total $^6$Li+$^7$Li abundance, $A$(Li), and the $^6$Li/$^7$Li
isotopic ratio, $q$(Li), which control line strength and line asymmetry, 
respectively, we also allow for a residual line broadening described by a 
Gaussian kernel with half-width $V_{\rm BR}$, and a global line shift, 
$\Delta v$. Note that the four fitting parameters are non-degenerate, since 
each one has a distinctly different effect on the line profile. The rotational 
line broadening is fixed to the value used in the 3D spectrum synthesis 
(we tried $v\,\sin i = 0$ and $2$~km/s). The value $q^\ast$(Li) of the best 
fit is then identified with the correction that has to be \emph{subtracted} 
from the $^6$Li/$^7$Li isotopic ratio derived from the 1D analysis to correct 
for the bias introduced by neglecting the intrinsic line asymmetry: 
$q^{(3D)}$(Li) = $q^{(1D)}$(Li) - $q^\ast$(Li). 
The procedure properly accounts for radiative transfer in the lines,
including saturation effects.

\begin{figure}
% \vspace*{-2.0 cm}
\begin{center}
\mbox{\includegraphics*[bb=20 28 580 370,width=0.96\textwidth]
{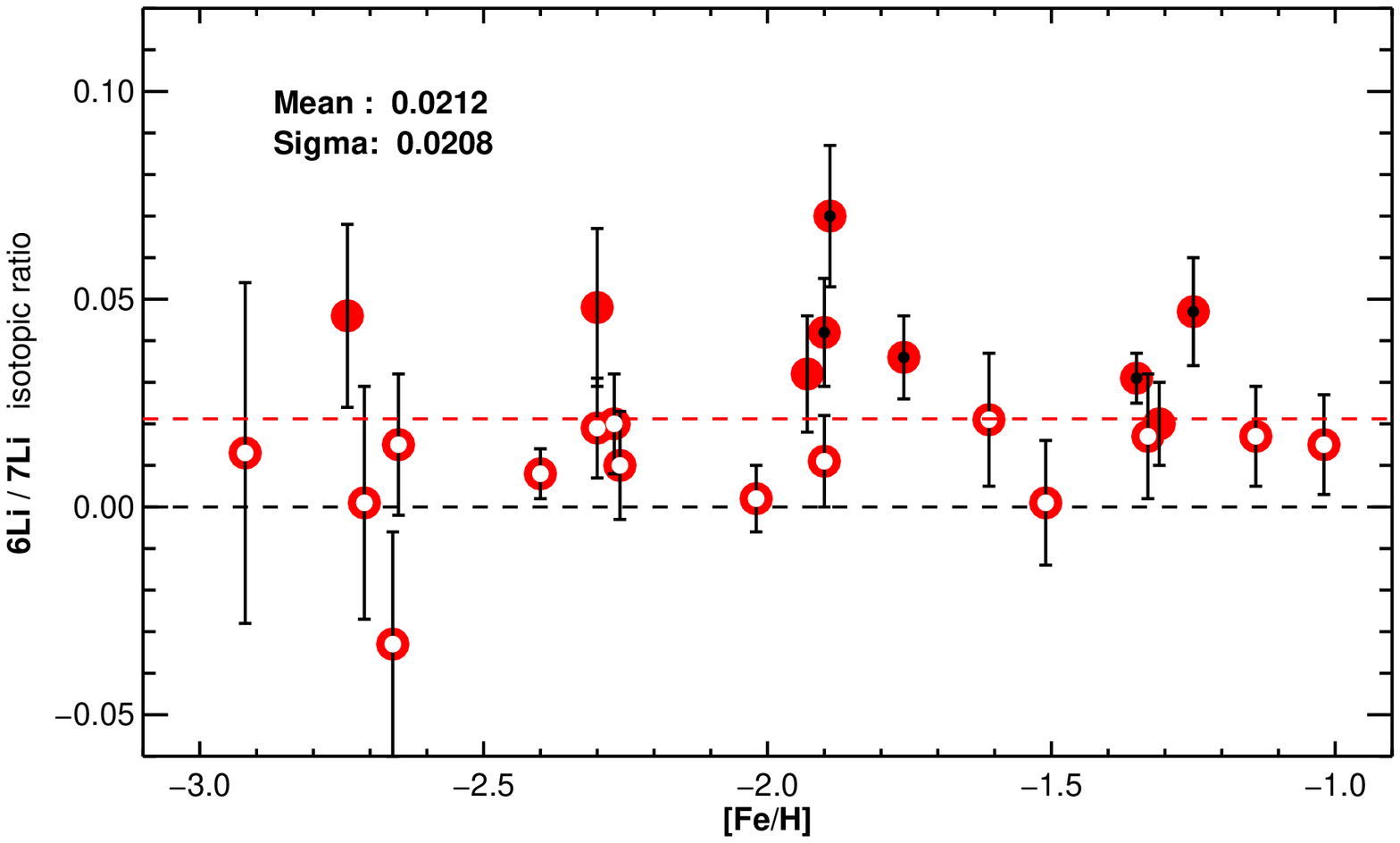}}
\mbox{\includegraphics*[bb=20 28 580 370,width=0.96\textwidth]
{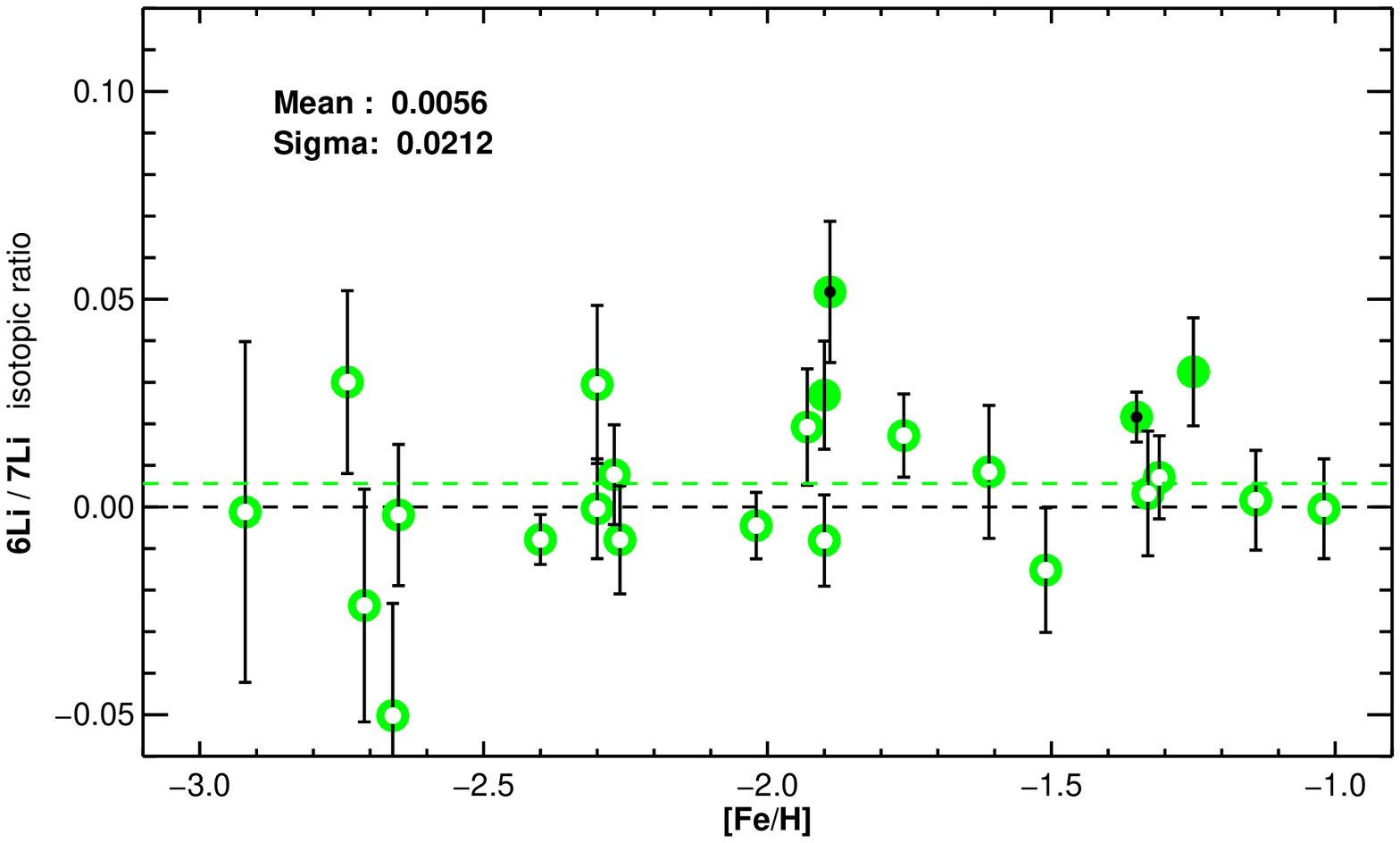}}
\vspace*{-0.1 cm}
\caption{$^6$Li/$^7$Li isotopic ratio, and $\pm 1 \sigma$ error bars, as a
function of metallicity as derived by \cite{A2006} before (top)
and after (bottom) subtraction of $q^\ast_b$(Li) (Table\,\ref{tab1}, col.~(9))
to correct for the bias due to the intrinsic line asymmetry. Open circles 
denote non-detections, filled circles and filled circles with a black dot 
denote $^6$Li detections above the $2\sigma$ and $3\sigma$ level, respectively.}
\label{fig2}
\end{center}
\end{figure}

Two different sets of 1D profiles were used for this purpose: (a) NLTE
line profiles based on the $\langle$3D$\rangle$ model, constructed by 
averaging the 3D model on surfaces of constant optical depth, and (b) 
LTE line profiles computed from a so-called LHD model, a 1D mixing-length model
atmosphere that has the same stellar parameters and uses the same microphysics
and radiative transfer scheme as the corresponding 3D model.
For both kind of 1D models, the microturbulence was fixed at 
$\xi_{\rm mic}=1.5$~km/s. The mixing length parameter adopted for the LHD 
models is $\alpha_{\rm MLT}=0.5$.

We have determined $q^\ast$(Li) according to the method outlined above
for our grid of 3D model atmospheres. The results are given in 
Table\,\ref{tab1}, both for fitting with the $\langle$3D$\rangle$~NLTE 
profiles ($q^\ast_a$(Li), col.\,(8)) and with the 1D~LTE profiles of
the LHD models ($q^\ast_b$(Li), col.\,(9)). At given
metallicity, the corrections are largest for low gravity and high
effective temperature. They increase towards higher metallicity. 
We also note that $q^\ast$(Li) is essentially insensitive to the choice 
of the rotational broadening.  

The analysis of \cite{A2006} utilizes 1D LTE profiles computed from
MARCS model atmospheres. Hence, the correction $q^\ast_b$(Li), 
computed with the 1D~LTE profiles of the 1D~LHD models, should be applied
to their $^6$Li/$^7$Li isotopic ratios. The resulting downward corrections 
are typically in the range $1\% < q^\ast_b$(Li) $< 2\%$ for the 
stars of their sample (cf.\ \cite[Steffen \etal\ 2010]{steffen2010}, Fig.\,2).
After subtracting the individual $q^\ast_b$(Li) for each of these
stars, according to their $T_{\rm eff}$, $\log g$, and [Fe/H], 
the mean $^6$Li/$^7$Li isotopic ratio of the sample is
reduced from $0.0212$ to $0.0056$, as illustrated in Fig.\,\ref{fig2}.
Given the $^6$Li/$^7$Li isotopic ratios and their $1\sigma$ error bars
as determined by \cite{A2006}, the number of stars with a $^6$Li detection 
above the 2$\sigma$ and 3$\sigma$ level is 9 and 5, respectively, out of
24. After correction, the number of 2$\sigma$ and 3$\sigma$ detections
is reduced to 4 and 2, respectively. One of the stars meeting the 3$\sigma$ 
criterion after correction, HD\,106038, survives only because of its 
particularly small error bar of $\pm 0.006$. The other one is
G020-024, which shows the clearest evidence for the presence of $^6$Li
($q$(Li)~$=0.052 \pm 0.017$), while HD\,102200 remains the only clear
2$\sigma$ detection ($q$(Li)~$=0.033 \pm 0.013$). The spectra of these stars
should be reanalyzed with 3D NLTE line profiles.\\[-6mm]

\begin{table}[b]
  \vspace*{-2mm}
  \begin{center}
  \caption{Observed Li\,I $\lambda\,6707$~\AA\ spectra of three metal-poor
           turn-off stars}
  \label{tab2}
 {\scriptsize
  \begin{tabular}{|l|c|c|c|c|c|c|c|}\hline
{\bf Star} & {$T_{\rm eff}$~[K]} & {\bf $\log g$} & {\bf [Fe/H]} & 
{\bf R=$\lambda/\Delta\lambda$} &
{\bf S/N} & {\bf Instrument} & {\bf Reference} \\
\hline
HD\,74000  & $6203$ & $4.03$ & -2.05 & $120\,000$ & $600$ &
{\sf ESO3.6 / HARPS}  & {Cayrel \etal\ 2007} \\
G271$-$162 & $6330$ & $4.00$ & -2.25 & $110\,000$ & $600$ &
{\sf VLT / UVES}      & {Nissen \etal\ 2000} \\
HD\,84937  & $6300$ & $4.00$ & -2.30 & $100\,000$ & $630$ &
{\sf CFHT / GECKO}    & {Cayrel \etal\ 1999} \\
\hline
  \end{tabular}
  }
 \end{center}
\end{table}

\section{Analysis of observed spectra: 3D NLTE versus 1D LTE line fitting}
As a further exercise, we have fitted the observed Li\,I
$\lambda\,6707$~\AA\ spectra of three halo turn-off stars (see
Table\,\ref{tab2}), both with 3D~NLTE and 1D~LTE synthetic line
profiles. Since the parameters of these three stars are very similar,
the synthetic spectra were computed in all cases from hydrodynamical
model d3t6300g40mm20n01 (see Table\,\ref{tab1}) and the corresponding
1D~LHD model. The standard approach is to vary the four parameters
$A$(Li), $q$(Li), $V_{\rm BR}$, and $\Delta v$ to find the optimum
line profile fit, as described above (method~I).
The results are presented in Table\,\ref{tab3}. As expected, the 3D
analysis yields lower $^6$Li/$^7$Li isotopic ratios by about 1.7\%.
The numbers differ slightly from those given by \cite[Steffen \etal\
2010]{steffen2010} due to an upgrade of the line fitting procedure,
but the conclusions remain unchanged: HD\,74000 and G271$-$162 are
considered non-detections, while HD\,84937 remains a clear $^6$Li
detection with $q^{(\rm 3D-NLTE)}$(Li) $\approx 5\%$.

For completeness, Table\,\ref{tab3} also shows the results of fitting
the observed lines with 3D LTE synthetic profiles (not recommended).
The $^6$Li abundances obtained in this way lie between the 3D NLTE
 and the 1D LTE results. It is not obvious why fitting with the more 
asymmetric 3D LTE profiles gives higher $^6$Li than fitting with the
slightly more symmetric 3D NLTE profiles, but this is a robust result.

For HD\,74000, we also tried an alternative approach (method~II), 
where $V_{\rm BR}$ is determined from other spectral lines,
and is then fixed during the fitting of the Li\,I $\lambda\,6707$~\AA\ 
feature. For this purpose, we rely on a set of 6 clean Fe\,I lines
with similar excitation potential between $2.4$ and $2.6$~eV: 
the 5 lines of \cite[Cayrel \etal\ (2007)]{Ca2007}, Table\,1, plus
Fe\,I $\lambda\,6230.7$~\AA, $E_i=2.559$~eV. Their equivalent widths
range from $W \approx 13$ to $30$~m\AA, embracing the strength of the 
Li doublet. Fitting the 6 Fe\,I lines with 3D LTE synthetic line profiles,
we obtain $V_{\rm BR}$(Fe) = $3.05~(2.15) \pm 0.16~(0.23)$~km/s for 
$v\,\sin i = 0~(2)$~km/s, in close agreement with $V_{\rm BR}$(Li) 
inferred from the fitting of the Li\,I profile. This result may be 
taken as an indication that, in contrast to Li, the selected Fe\,I 
lines are not severely affected by departures from LTE. We note, 
however, a trend of $V_{\rm BR}$(Fe) increasing slightly with $W$, 
which remains to be understood. If instead the Fe\,I lines are fitted
with 1D LTE synthetic profiles, we find  $V_{\rm BR}$(Fe) = 
$5.61~(5.13) \pm 0.07~(0.09)$~km/s, systematically lower than 
$V_{\rm BR}$(Li) by $0.3$~km/s. Fixing $V_{\rm BR}$=$V_{\rm BR}$(Fe),
the best fit of the Li\,I doublet implies $q^{\rm 1D~LTE}$(Li)=$1.7$\%, 
which is $1.1$\% higher than with method~I. Finally, Table\,\ref{tab3} 
demonstrates that applying method~II with 3D LTE fitting of Li\,I leads
to a severe overestimation of the $^6$Li/$^7$Li isotopic ratio:
$q^{\rm 3D~LTE}$(Li) $>4$\% compared to $q^{\rm 3D~NLTE}$(Li)=$-1.1$\%!\\[-6mm]

\begin{table}
  \vspace*{-2mm}
  \begin{center}
  \caption{Fitting the observed spectra of Table\,\ref{tab2}
           with 3D and 1D synthetic line profiles.
           Columns (4)-(7) show the results for $v\,\sin i$ = 0.0 / 2.0 km/s.}
  \label{tab3}
 {\scriptsize
  \begin{tabular}{|l|c|c|c|c|c|c|}\hline
{\bf Star}            & {\bf synthetic} & {\bf fitting} & 
{\bf $A$(Li)}$^{~1)}$ & {\bf n($^6$Li)/n($^7$Li)} & {\bf $V_{\rm BR}^{~2)}$} & 
{\bf $\Delta v$} \\
{}                    & {\bf spectrum}  & {\bf method}  &
                      & {\bf [\%]}      & {\bf [km/s]}  & 
{\bf [km/s]}     \\
\hline
HD\,74000  & 3D NLTE &  I  &
$2.25$ / $2.25$ & {\bf -1.1 / -1.1} & $3.1$ / $2.1$ & ~0.64 / ~0.64 \\
 & (3D~~~LTE)        &  I  &
$1.85$ / $1.85$ & {\bf -0.4 / -0.5} & $4.7$ / $4.1$ & ~0.66 / ~0.67 \\
 & 1D~~~LTE          &  I  &
$2.23$ / $2.23$ & {\bf ~0.6 / ~0.6} & $5.9$ / $5.4$ & ~0.43 / ~0.43 \\
           & 3D NLTE & II  &
$2.25$ / $2.25$ & {\bf -1.1 / -1.3} & {\bf 3.1 / 2.2} & ~0.64 / ~0.65 \\
 & (3D~~~LTE)        & II  &
$1.84$ / $1.84$ & {\bf ~4.6 / ~4.2} & {\bf 3.1 / 2.2} & ~0.46 / ~0.47 \\
 & 1D~~~LTE          & II  &
$2.23$ / $2.23$ & {\bf ~1.7 / ~1.7} & {\bf 5.6 / 5.1} & ~0.39 / ~0.39 \\
G271$-$162 & 3D NLTE &  I  &
$2.30$ / $2.30$ & {\bf ~0.7 / ~0.6} & $3.7$ / $2.9$ & ~0.04 / ~0.04 \\
 & (3D~~~LTE)        &  I  &
$1.89$ / $1.89$ & {\bf ~1.3 / ~1.1} & $5.2$ / $4.6$ & ~0.06 / ~0.07 \\
 & 1D~~~LTE          &  I  &
$2.27$ / $2.27$ & {\bf ~2.3 / ~2.4} & $6.2$ / $5.8$ & -0.17 / -0.17 \\
HD\,84937  & 3D NLTE &  I  &
$2.20$ / $2.20$ & {\bf ~5.2 / ~5.2} & $3.3$ / $2.4$ & ~0.02 / ~0.02 \\
 & (3D~~~LTE)        &  I  &
$1.80$ / $1.80$ & {\bf ~5.8 / ~5.7} & $4.8$ / $4.3$ & ~0.05 / ~0.05 \\
 & 1D~~~LTE          &  I  &
$2.18$ / $2.18$ & {\bf ~6.9 / ~6.9} & $6.0$ / $5.6$ & -0.19 / -0.19 \\
\hline
  \end{tabular}
  }
 \end{center}
\vspace{1mm}
 \scriptsize{
 {\it Notes:}
  $^{1)}$ $\log \left[n(^6{\rm Li})+n(^7{\rm Li})\right] -
  \log n({\rm H)}+12$; $^{2)}$ Gaussian kernel, {\bf bold:} fixed from 
   Fe\,I lines}
\end{table}

\section{Conclusions}
The $^6$Li/$^7$Li isotopic ratio derived from fitting of the Li\,I doublet 
with 3D NLTE synthetic line profiles is shown to be about 1\% to 
2\% lower than what is obtained with 1D LTE profiles. 
Based on this result, we conclude that only $2$ out of the
$24$ stars of the \cite{A2006} sample would remain significant $^6$Li
detections when subjected to a 3D~non-LTE analysis, suggesting that
the presence of $^6$Li in the atmospheres of galactic halo stars is
rather the exception than the rule, and hence does not necessarily
constitute a \emph{cosmological} $^6$Li problem.
If we adopt the approach 
by \cite{A2006}, relying on additional spectral lines to fix the 
residual line broadening, the difference between 3D NLTE and 1D LTE results 
increases even more, as far as we can judge from our case study HD\,74000.
Until the 3D NLTE effects are fully understood for all involved lines, 
we consider this method as potentially dangerous.\\[-6mm]

%\begin{discussion}
%
%\end{discussion}

\end{document}